\documentclass[a4paper,11pt]{article}
\pdfoutput=1 

\usepackage{jinstpub} 
\usepackage{epstopdf}
\usepackage{bm}
\usepackage{subfigure}
\title{\boldmath Performance of Linear Classification Algorithms on $\alpha/\gamma$ Discrimination for LaBr3:Ce Scintillation Detectors with Various Pulse Digitizer Properties}

\author[a,b]{Jingjun Wen,}
\author[a,b]{Jinfu Zhu,}
\author[a,b,1]{Tao Xue,\note{Corresponding author.}}
\author[a,b,c]{Jirong Cang,}
\author[a,b,d]{Liangjun Wei,}
\author[a,b]{Qiyuan Nie,}
\author[a,b]{Ming Zeng,}
\author[a,b]{Zhi Zeng,}
\author[a,b]{Hao Ma}
\author[a,b,d]{Jianmin Li,}
\author[a,b]{and Yinong Liu,}

\affiliation[a]{Key Laboratory of Particle \& Radiation Imaging (Tsinghua University), Ministry of Education,\\Beijing, China}
\affiliation[b]{Department of Engineering Physics, Tsinghua University,\\Beijing, China}
\affiliation[c]{Department of Astronomy, Tsinghua University,\\Beijing, China}
\affiliation[d]{NUCTECH Company Limited,\\Beijing, China}

\emailAdd{xuetaothu@mail.tsinghua.edu.cn}

\abstract{With the development of high-speed readout electronics, the digital pulse shape discrimination (PSD) methods have attracted the attention of more researchers, especially in the field of high energy physics and neutron detection. How to choose a PSD algorithm and corresponding data acquisition system (DAQ)  naturally becomes a critical problem to settle down for the detection system designers. In this paper, the relationship between the classification performance of different PSD algorithms and digitizers' sampling properties (including sampling rate and the effective number of bits) has been researched based on $\mathrm{LaBr_3}$:Ce scintillation detectors. A self-developed integrated digitizer configured with five different ADCs and a WavePro 404HD oscilloscope were deployed to digitize the waveforms from $\mathrm{LaBr_3}$:Ce scintillators. Moreover, three PSD methods, charge comparison method (CCM), least square for classification method (LS) and Fisher's linear discriminant analysis (LDA), based on linear model were applied to discriminate the alpha signals from the intrinsic isotope $^{227}$Ac. With the LS method and a 125 MSPS 14-Bit ADC, the FoM value was 1.424$\pm$0.042, which is similar to the result from LDA but 31\% better than the result of CCM. The discrimination results showed that the performances of LS and LDA are less affected by the sampling rate with respect to the CCM method, which reflects in a better PSD capability. The results of this paper can help the developers of detector systems to make a trade-off among sampling properties, desirable discrimination results and the cost of systems.}


\begin{document}
	\maketitle
	\flushbottom
	
	\section{Introduction}
	\label{sec:intro}
	$\mathrm{LaBr_3}$:Ce scintillation detectors are widely used in energy and TOF measurement for their excellent energy resolution (< 3\% at 662 keV ) \cite{dorenbos2002light} and time resolution ($\sim$ 300 ps). However, due to the contribution of the five short-lived progeny of an Ac impurity, the $\alpha$ background with energy region between 1,500 keV and 2,750 keV in $\mathrm{LaBr_3}$:Ce scintillators cannot be ignored. It has a strong impact on the application of this detector in the low-activity $\gamma$ measurement. Discriminating against the $\alpha$ background from $\gamma$-ray signals effectively is very important, and the digitizer system associated with pulse shape discrimination (PSD) is on demand.
	
	Digital PSD methods for $\alpha/\gamma$ has been extensively studied and improved owing to the development of fast and high precision digitizer systems in the past decade. How to make a trade-off between digitizer's sampling properties and discrimination efficiency has been a critical problem. Recently, several research has investigated the relationship between the PSD performance of scintillators and the sampling properties of digitizers. M. Flaska and D. Cester studied the influence of sampling properties on $\mathrm{n}/\gamma$ discrimination for organic scintillation detectors and verified that the oversampling could improve discrimination efficiency \cite{RN24,RN22,ref4}. J. Zhang et al. focused on the influence of sampling rate on $\mathrm{n}/\gamma$ discrimination for stilbene scintillation detectors and researched the improvement caused by digital signal process (DSP) algorithms \cite{RN25}. For $\mathrm{LaBr_3}$:Ce scintillators, J. Cang et al. studied the influence of sampling properties on the energy resolution and $\alpha/\gamma$ discrimination for $\mathrm{LaBr_3}$:Ce and proposed a quantitative model for optimal digitizer selection based on the Charge Comparison Method (CCM)\cite{RN27}. Although the above studies have compared the influence of sampling properties on PSD performance, they were limited to the use of CCM alone. In this paper, we analyze the distribution of $\mathrm{LaBr_3}$:Ce scintillation detector's data and then describe the study on the performance of three linear PSD algorithms with different sampling properties. 
	
	In this paper, we show the results on the analysis performed using the t-Distributed Stochastic Neighbor Embedding (t-SNE) algorithm to get the qualitative information on the waveform distribution from experimental data. These results illustrated that the classification algorithm based on linear model is effective enough for the $\alpha/\gamma$ discrimination on $\mathrm{LaBr_3}$:Ce detector. From this perspective, three PSD algorithms based on linear models (CCM, LS and LDA) were evaluated in this paper, and their performances have been compared in the same experimental situation.
	
	Besides, the relationship between the performances of PSD algorithms and the sampling properties (including sampling rates, effective number of bits and input ranges) of digitizers have been intensively studied. A self-developed digitizer system with various ADCs and a LeCroy WavePro 404HD oscilloscope were deployed in the experiment to provide an extensive range of sampling rates (from 100 MSPS to 5 GSPS) and effective number of bits (from 7.5 to $\sim$12). According to the experimental results, the relationships between CCM's performances and sampling properties have been derived from the ADC quantization noise theory, and the other two algorithms based on linear model were discussed qualitatively. This paper also refers on the study of the influence of the minimal differences between $\alpha/\gamma$ tail waveforms on PSD results and discuss why digitizers' effective number of bits (ENOB) play a critical role in the discrimination. The results shown in this paper may help detectors system developers to make a trade-off or balance between DAQ's sampling properties and desirable discrimination results, and provide preliminary guidance for the design of a digital real-time PSD system.
	
	The rest of this paper is organized as follows: Section 2 describes the experimental setup and the corresponding hardware's properties, Section 3 introduces three PSD algorithms based on linear model: CCM, LS and LDA and explains why such algorithms can provide an optimum discrimination efficiency in $\mathrm{LaBr_3}$:Ce $\alpha/\gamma$ classification. Section 4 shows the data processing results and analyzes the relationship among PSD performances, sampling properties and sampling intervals, Section 5 compares the PSD capabilities of the three algorithms and discusses the advantages of such linear algorithms, Section 6 concludes the paper.
	
	\section{Experiment Setup}
	\label{sec:expset}
	\subsection{Detector System}

\begin{figure}[htbp]
		
		\centering 
		\includegraphics[width=1.0\textwidth,clip]{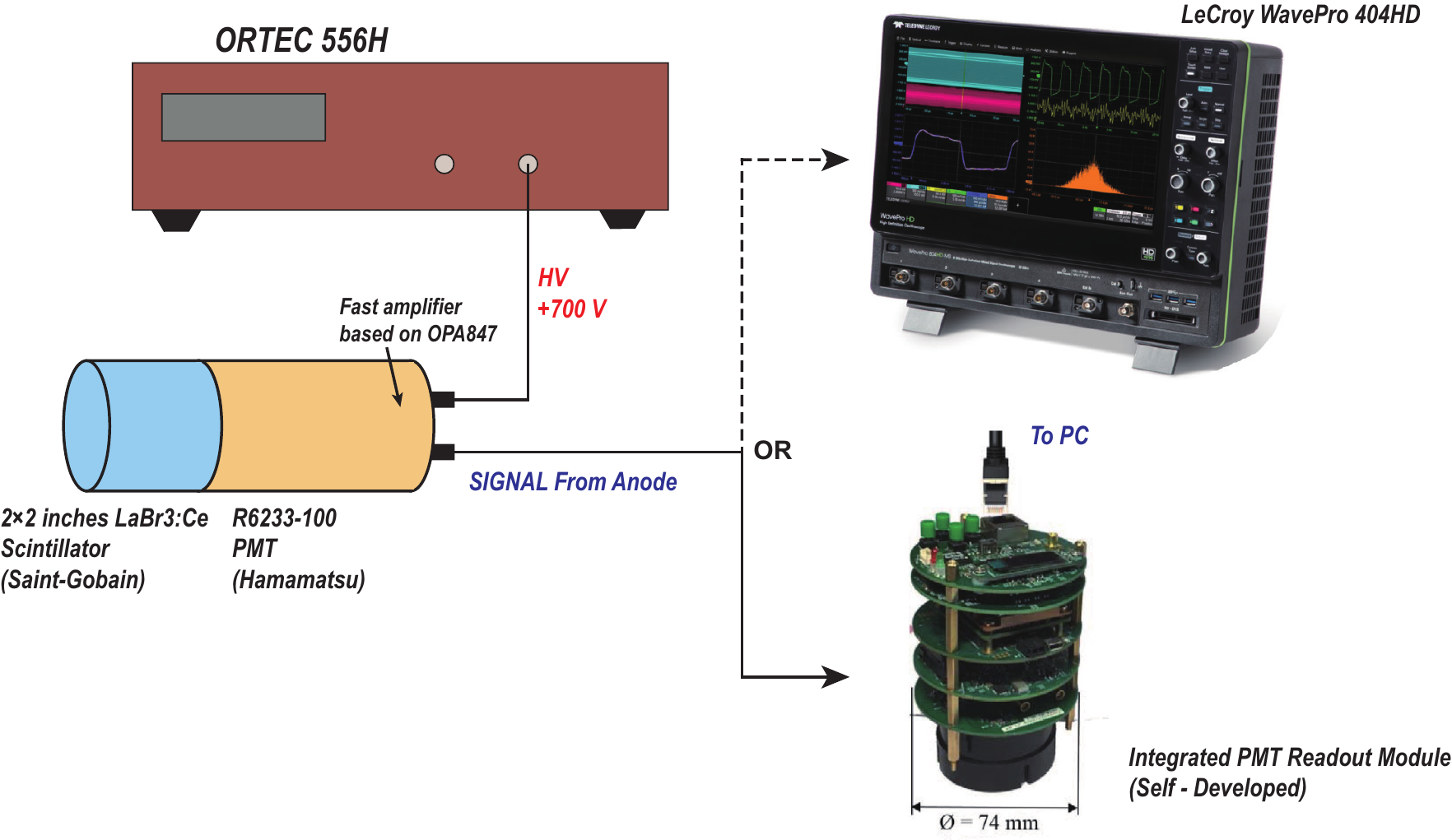}
		\caption{\label{fig:1} Experiment setup of this research.}
	\end{figure}

	The experiment setup is shown in figure \ref{fig:1}, a $2^{''}\times 2^{''}$ cylindrically shaped $\mathrm{LaBr_3}$:Ce scintillation detector  (Saint-Gobain BrilLanCe$^\text{TM}$ 380) coupled with a Hamamatsu R6233-100 photomultiplier tube is used to obtain the signal generated by particles. The ORTEC 556H high voltage module supplies +700 V high voltage to the PMT, and a fast amplifier is directly connected to the anode of PMT, and the amplified signals (voltage gain is $~$7) is directly digitized by the digitizer systems. The integrated PMT digitizer system or a LeCroy WavePro 404HD oscilloscope is deployed as a DAQ system for the full digitization of every pulse shape. The sampling properties, including sampling rate, vertical resolution, ENOB and input range, are listed in Table \ref{tab:1}. 
	
	The experiment without any radioactive source and shield was designed to obtain the $\gamma$ signals from background and $\alpha$ signals from intrinsic isotopes in the meantime. Each test has recorded over 1,100,000 particle events with energy between 1600 keV and 2800 keV.
	
	\subsection{Digitizers \& Sampling Properties}
	\begin{table}[htbp]
		\centering
		\caption{\label{tab:1} Samping properties of the integrated PMT digitizers and the LeCroy WavePro 404HD  oscilloscope}
		\smallskip
		\begin{tabular}{|l|c|c|c|c|c|}
			\hline
			ADC&\hspace{0.865em}AD9233\hspace{0.865em} & \hspace{0.865em}AD9255\hspace{0.865em} &\hspace{0.865em}AD9265\hspace{0.865em}  &\hspace{0.865em}AD9642\hspace{0.865em} &ISLA212P50\\ 		
			\hline
			Sampling Rate & 125 MSPS & 125 MSPS & 125 MSPS & 250 MSPS & 500 MSPS\\
			Vertical Resolution & 12-Bit & 14-Bit & 16-Bit & 14-Bit & 12-Bit\\
			ENOB & 11.18 & 11.97 & 11.87 & 11.06 & 10.52\\
			Full Scale & 2000 mV & 2000 mV & 2000 mV & 1750 mV & 2000 mV\\
			$[FullScale/(\sqrt{Fs}\cdot 2^{\mathrm{ENOB}})]^2$ & $5.94\times10^{-9}$ & $1.99\times10^{-9}$ & $2.28\times10^{-9}$ & $2.69\times10^{-9}$ & $3.71\times10^{-9}$  \\
			\hline
		\end{tabular}
		\smallskip
		\begin{tabular}{|l|c|c|c|c|c|}
			\hline
			ADC&Oscilloscope&Oscilloscope&Oscilloscope&Oscilloscope&Oscilloscope\\ 		
			\hline
			Sampling Rate & 100 MSPS & 500 MSPS & 1 GSPS & 2.5 GSPS & 5 GSPS\\
			Vertical Resolution & 12-Bit & 12-Bit & 12-Bit & 12-Bit & 12-Bit\\
			ENOB & 7.5 & 7.5 & 7.5 & 7.5 & 7.5\\
			Full Scale & 1600 mV & 1600 mV & 1600 mV & 1600 mV & 1600 mV\\
			$[FullScale/(\sqrt{Fs}\cdot 2^{\mathrm{ENOB}})]^2$ & $7.81\times10^{-7}$ & $1.56\times10^{-7}$ & $7.81\times10^{-8}$ & $3.13\times10^{-8}$ & $1.56\times10^{-8}$  \\
			\hline
		\end{tabular}
	\end{table}
	The integrated PMT digitizer system used in the experiment can be configured with various ADCs whose sampling rates from 125 MSPS to 500 MSPS and ENOB from 10.5 to 12 \cite{RN28}. Moreover, a LeCroy WavePro 404HD oscilloscope was also used to provide a much higher sampling rate (up to 5 GSPS) but lower ENOB (7.5). 
	
	\section{Algorithms}
	\subsection{Preparation of the Dataset}
\begin{figure}[htbp]
		\centering 
		\includegraphics[width=.8\textwidth,clip]{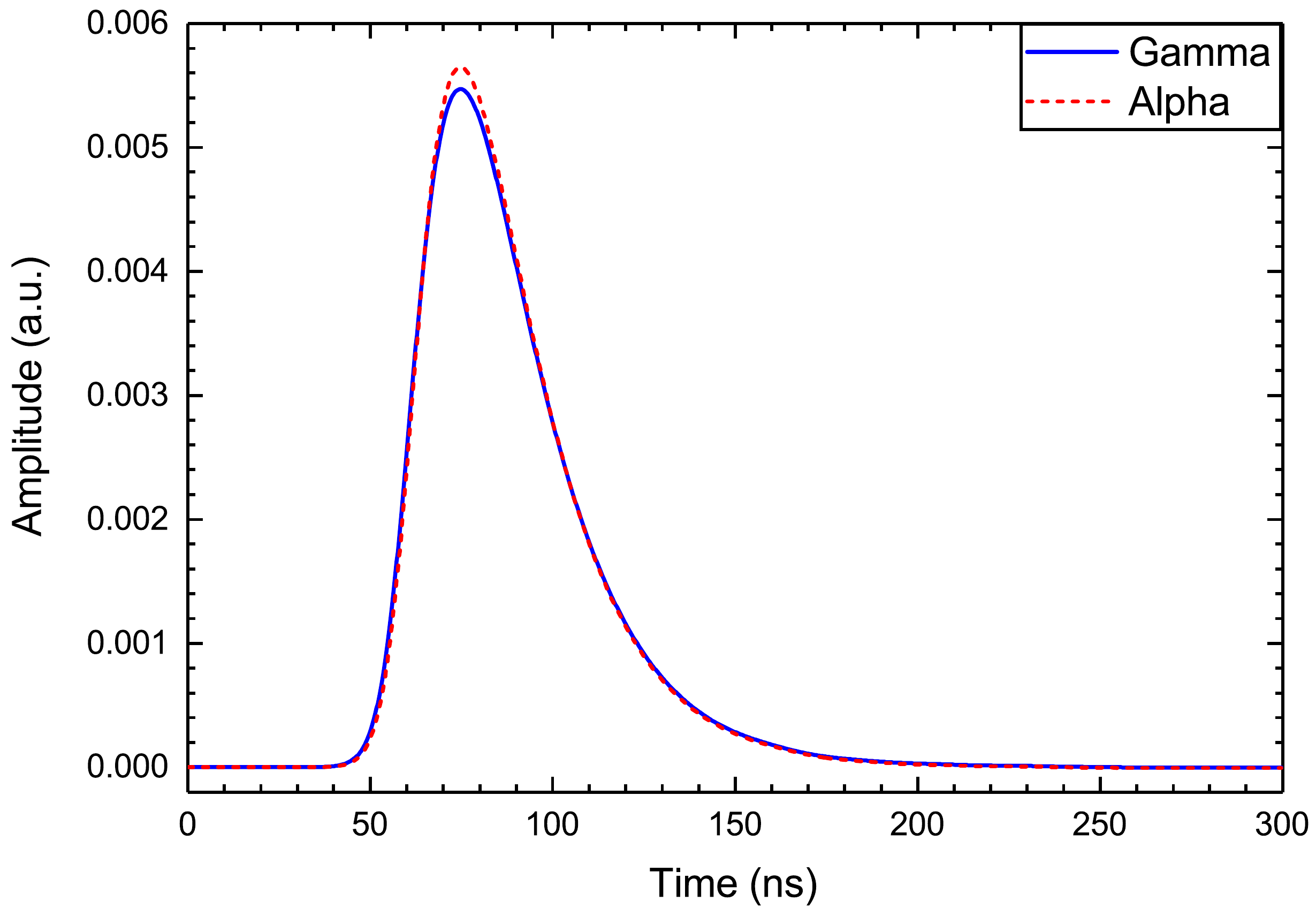}
		\caption{\label{fig:2} Average $\alpha/\gamma$ waveforms measured by a 500 MSPS, 12-Bit ISLA212P50 ADC with $\mathrm{LaBr_3}$:Ce detector. Each pulse shape is the mean result of 50,000 pulses, and normalized by pulse area which is proportional to the energy of the particle.}
	\end{figure}

	The average $\alpha/\gamma$ digitized signals of $\mathrm{LaBr_3}$:Ce scintillator from the self-developed integrated PMT digitizer are shown in figure \ref{fig:2}.
	In order to minimize the influence of threshold jitter and high-frequency electronic noise, some necessary preprocess shown below were applied to the $\mathrm{LaBr_3}$:Ce dataset:
	\begin{enumerate}
	    \item The average value of the raw data ranging from 0$\sim$30 ns is selected to subtract the baseline of the total waveform.
		\item The waveforms registered with a sampling rate lower than 5 GSPS were interpolated to 5 GSPS by the spline interpolation to guarantee accurate alignment performance in the next process step.
		\item The third-order Butterworth low pass filter with 50 MHz cutoff frequency was used to suppress the high-frequency electronic noise. It should be emphasized that the Butterworth filter did not change the envelope of the waveform but remove the high-frequency noise.
		\item Every waveform was aligned to the location of the peak value (75.0 ns) to achieve accurate alignment.  
	\end{enumerate}
	After the preprocess, the energy of every pulse was calibrated by the intrinsic $\gamma$ radioactivity, and the $\gamma$ waveforms with energy between 1600 and 2800 keV and $\alpha$ waveforms with the same integral charge were selected for $\alpha/\gamma$ discrimination dataset. 
	
	\subsection{t-Distributed Stochastic Neighbor Embedding (t-SNE) Algorithm}
	\label{sec:3.2}
	\begin{figure}[htbp]
		\centering 
		\includegraphics[width=\textwidth,clip]{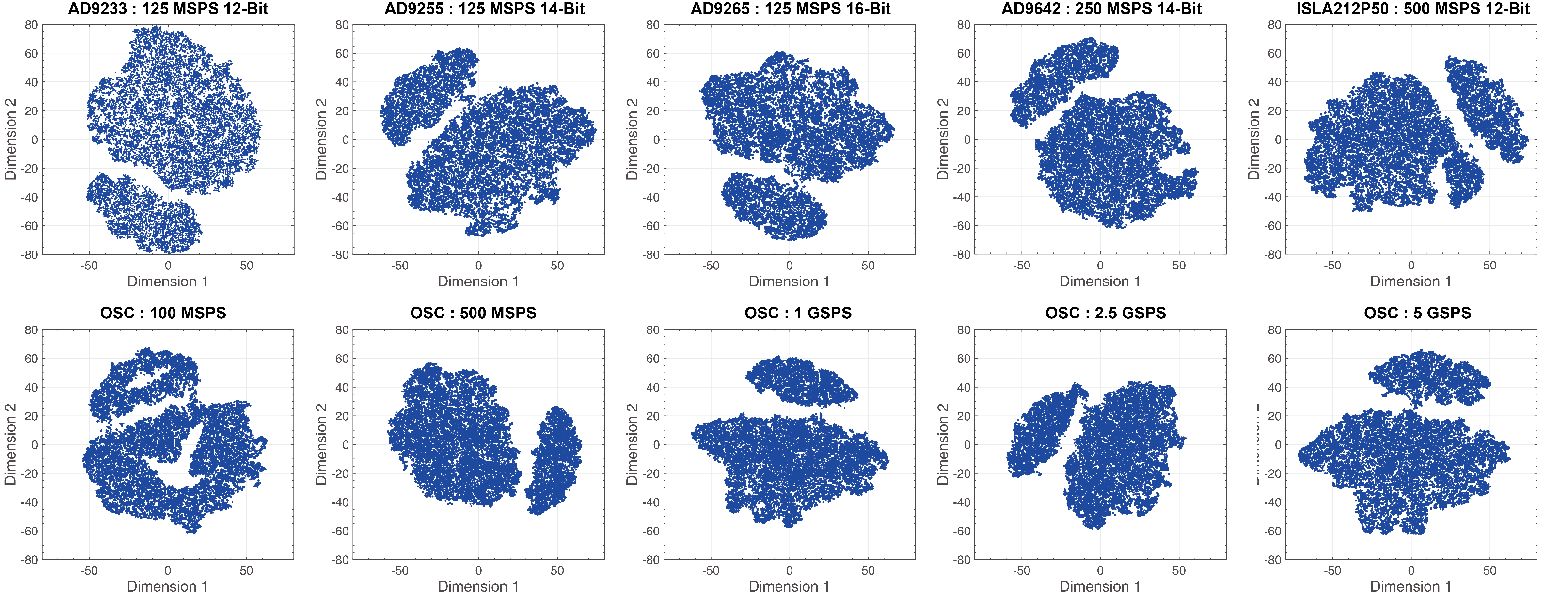}
		\caption{\label{fig:3} tSNE results of all datasets from $\mathrm{LaBr_3}$:Ce.}
	\end{figure}
	t-SNE is a nonlinear dimensionality reduction algorithm for embedding high-dimensional data for visualization in a low-dimensional space. The main idea of t-SNE is converting Euclidean distance to conditional probability to express the similarity among data points, that is to say, the t-SNE maps data points to a probability distribution through an affine transformation, which mainly includes three steps:
	\begin{itemize}
		\item t-SNE constructs a probability distribution among high-dimensional data, which makes similar data points have higher probability to be selected, while different data points have lower probability to be selected. Given a dataset $\mathbf{D} = \{\bm{x}_1,\bm{x}_2,\cdots,\bm{x}_n\},\bm{x}_i = (x_{i1};\cdots;x_{id})$, t-SNE first calculates the joint probability $p_{ij}$, which is proportional to the similarity between $\bm{x}_i$ and $\bm{x}_j$. In order to obtain the $p_{ij}$, the conditional probability $p(i|j)$ is calculated by equation \ref{eq:sub1}.
		\begin{equation}\label{eq:sub1}
		p(i|j) = \dfrac{\exp\left(-||\bm{x}_i-\bm{x}_j||^2/2\sigma_i^2\right)}{\displaystyle\sum_{k\neq i}\exp\left(-||\bm{x}_i-\bm{x}_k||^2/2\sigma_i^2\right)}
		\end{equation}
		where $\sigma_i$ is the variance of the Gaussian that is centered on data point $\bm{x}_i$. And then
		\begin{equation}
		p_{ij} = \dfrac{p(i|j)+p(j|i)}{2n}
		\end{equation}
		\item t-SNE constructs the probability distribution (Student t-distribution) of these data points in low dimensional space (2-D or 3-D), and guarantee those two probability distributions (in high-dimensional space and low-dimensional space) are as similar as possible. For the low-dimensional counterparts $\bm{l}_i$ and $\bm{l}_j$ of the high-dimensional data points $\bm{x}_i$ and $\bm{x}_j$, the joint probability $q_{ij}$ can be calculated by equation \ref{eq:sub2}.
		\begin{equation}\label{eq:sub2}
		q_{ij} = \dfrac{\left(1+||\bm{l}_i-\bm{l}_j||^2\right)^{-1}}{\displaystyle\sum_{m\neq n}\left(1+||\bm{l}_m-\bm{l}_n||^2\right)^{-1}}
		\end{equation}
		\item In order to make $q_{ij} = p_{ij}$, t-SNE algorithm minimizes the sum of Kullback-Leibler divergences over all data points using a gradient descent method. The cost function $\mathcal{C}$ is given by equation \ref{eq:sub3}.
		\begin{equation}\label{eq:sub3}
		\mathcal{C} = \displaystyle\sum_iKL(P_i||Q_i) = \displaystyle\sum_i\sum_jp_{ij}\log\dfrac{p_{ij}}{q_{ij}}
		\end{equation}
	\end{itemize}
	
	In this paper, t-SNE algorithm was used to gain the dimension reduction results of datasets measured by different digitizers. The results are shown in figure 3, dimension 1 and dimension 2 are two dimensions of $\bm{l}_i$ (i.e. x-axis and y-axis in 2-D). As portrayed in figure 3, the results represent two features that are linearly separable in two dimensions. Based on this dimensionality reduction results, combined with the theory of signal generation mechanism of scintillation detectors, it can be assumed that $\alpha/\gamma$ waveforms of $\mathrm{LaBr_3}$:Ce detectors are linearly separable.
	
	\subsection{PSD Algorithms}
	In this section, we will introduce three PSD algorithms based on linear model: CCM, LS and LDA. The algorithms are characterized by using linear transformation to calculate the classification features ($y=\bm{w}^T\bm{x}+b$) and provide optimum performance on the discrimination of linearly separable datasets, where $\bm{w}$, $b$ are respectively the weight vector and bias ($\bm{w}^T$ is the transposed vector of $\bm{w}$).
	\subsubsection{CCM}
	The CCM is widely used in the PSD for its sufficient discrimination efficiency, ease of use and robust performance. It's based on a comparison of the integrals of waveforms, over two different intervals called long integral and short integral. The long integral generally refers to the integral of the entire pulse, while the short integral includes part of the pulse area where the difference is larger. Naturally, we usually integrate in the short interval choosing the one in which the major difference between the signals coming from $\alpha$ and $\gamma$ is observed. The test statistic of CCM can be expressed as:
	\begin{equation}
	\label{eq:1}
	CCM = Q_S/Q_L = \dfrac{\displaystyle\sum_{t=t_1}^{t_2}x(t)}{\displaystyle\sum_{t_{total}}x(t)}
	\end{equation}
	where $Q_S$ represents the short integral and $Q_L$ is the long integral. In this paper, the lower bound of the short integral is $t_1=62.8$ ns and the upper bound of it is $t_2=103$ ns \cite{RN29}, which ensures the $\alpha/\gamma$ data are separated to the greatest extent. 
	
	Since CCM has no demand for the labeled $\alpha/\gamma$ waveforms, its discrimination results can provide the training datasets for other supervised PSD algorithms such as LDA, LS and Neural Network.
	\subsubsection{Least Square Method for classification (LS)}
	LS is a supervised PSD method which calculates the weight $\bm{w} = (w_1;w_2;\cdots;w_n)$ by minimizing a sum-of-squares error function of the training dataset \cite{bishop2006pattern}. Given a training dataset 
	
	\noindent{$\mathbf{D} = \{(\bm{x}_1,y_1),(\bm{x}_2,y_2),\cdots,(\bm{x}_n,y_n)\}$, where $\bm{x}_i = (x_{i1};x_{i2};\cdots;x_{id})$ is the waveform of $\mathrm{LaBr_3}$:Ce and $y_i\in\{-1,+1\}$ represents if the waveform is generated by $\alpha$ particles ($y_i=+1$) or $\gamma$-rays ($y_i=-1$). The target of LS is to find a weight vector $\bm{w}$ and bias coefficient $b$, such that $\bm{w}^T\bm{x}_i+b\approx y_i$. }
	
	Let
	\begin{equation}
	\label{eq:2}
	\hat{\bm{w}} = (\bm{w};b)^T,\quad \mathbf{X} = \begin{bmatrix}
	\bm{x}_1^T & 1\\
	\vdots & \vdots\\
	\bm{x}_n^T & 1
	\end{bmatrix},
	\quad \bm{y} = (y_1,y_2,\cdots,y_n)^T
	\end{equation} 
	Using the least square method to obtain the parameter estimation, we can write the target function as:
	\begin{equation}
	\label{eq:3}
	\hat{\bm{w}}^*=\mathop{\text{arg~min}}\limits_{\hat{\bm{w}}}(\bm{y}-\mathbf{X}\hat{\bm{w}})^T(\bm{y}-\mathbf{X}\hat{\bm{w}})
	\end{equation}
	The closed-form of the optimal solution of  $\hat{\bm{w}}$ can be solved by deriving equation \ref{eq:3} and making it zero.
	\begin{equation}
	\label{eq:4}
	\hat{\bm{w}}^* = \left(\mathbf{X}^T\mathbf{X}\right)^{-1}\mathbf{X}^T\bm{y}
	\end{equation} 
	In this research, the $\left(\mathbf{X}^T\mathbf{X}\right)^{-1}$ is the generalized inverse matrix of $\left(\mathbf{X}^T\mathbf{X}\right)$, because the matrix is singular.
	
\begin{figure}[htbp]
		\centering
		\includegraphics[width=\textwidth]{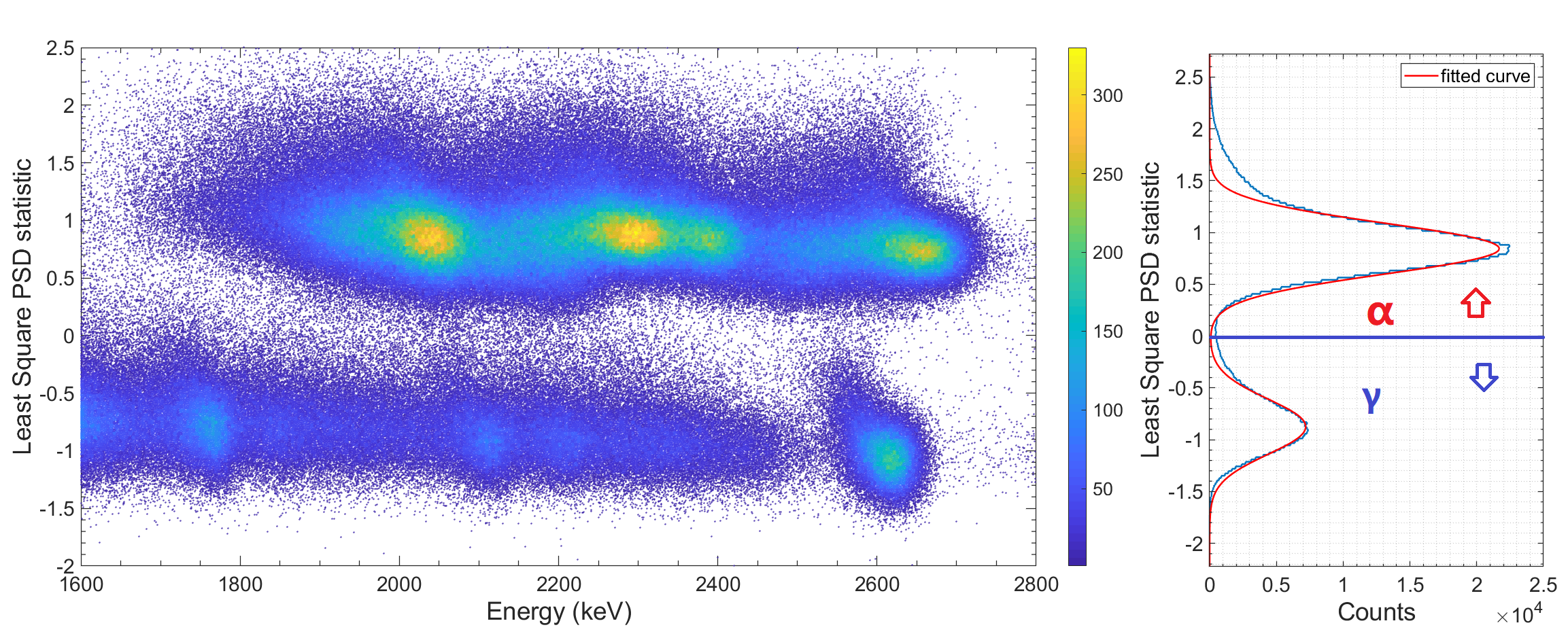}
		\caption{2D distribution of LS test statistic and energy measured by a 125 MSPS 14-Bit AD9255 ADC and the figure includes 1,000,000 particle events.\label{fig:LS}}
	\end{figure}

	Once the weight vector $\hat{\bm{w}} = (\bm{w},b)$" is obtained by the training data, the LS test statistics can be calculated by $y_i = \bm{w}^T\bm{x}_i+b$. The 2D distribution of LS test statistic and energy measured by AD9255 ADC (125 MSPS, 14-Bit) is shown in figure \ref{fig:LS}. The statistics of $\alpha$ particles and $\gamma$-rays are individually concentrated at +1 and -1, but the mean values of them are slightly shifting. So the final determination of whether the signal is $\alpha$ or $\gamma$ needs further study on the distribution of LS test statistics.

	\subsubsection{Fisher Linear Discriminant Analysis (LDA)}
	LDA is a classic linear classification algorithm that was first proposed by R. A. Fisher to solve the two-classes classification problem \cite{fisher1936use}. The main idea of this method is to find a projection direction that maximizes the separation between the projected classes' mean values while minimizes the variance within each class. 
	
	Given two training datasets $\mathbf{D}_\alpha = \{\bm{x}_{\alpha1},\bm{x}_{\alpha2},\cdots,\bm{x}_{\alpha n}\}$, $\mathbf{D}_\gamma = \{\bm{x}_{\gamma1},\bm{x}_{\gamma2},\cdots,\bm{x}_{\gamma n}\}$, where $\bm{x}_{\alpha i}/\bm{x}_{\gamma i}$ are the waveforms of $\alpha/\gamma$. Then, define the mean vectors and covariance matrix of $\mathbf{D}_\alpha/\mathbf{D}_\gamma$ as $\bm{\mu}_\alpha/\bm{\mu}_\gamma$ and $\bm{\Sigma}_\alpha/\bm{\Sigma}_\gamma$. The covariance matrix $\bm{\Sigma}_\alpha/\bm{\Sigma}_\gamma$ can be calculated by equation \ref{eq:Sigma}
	\begin{equation}\label{eq:Sigma}
	\bm{\Sigma_\alpha} = \sum_{\bm{x}_i\in\alpha}(\bm{x}_i-\bm{\mu_\alpha})(\bm{x}_i-\bm{\mu_\alpha})^T\qquad 	\bm{\Sigma_\gamma} = \sum_{\bm{x}_i\in\gamma}(\bm{x}_i-\bm{\mu_\gamma})(\bm{x}_i-\bm{\mu_\gamma})^T
	\end{equation}
	So the separation of projected classes' mean values can be expressed as $\left|\left|\bm{w}^T\bm{\mu}_\alpha - \bm{w}^T\bm{\mu}_\gamma \right|\right|_2^2$, and the sum of variances of the two classes is $\bm{w}^T\bm{\Sigma}_\alpha\bm{w}+\bm{w}^T\bm{\Sigma}_\gamma\bm{w}$. According to the principle of LDA, the target function is:
	
	\begin{equation}
	\label{eq:5}
	\bm{w}^* =\mathop{\text{arg~max}}\limits_{\bm{w}}\dfrac{\bm{w}^T(\bm{\mu}_\alpha-\bm{\mu}_\gamma)(\bm{\mu}_\alpha-\bm{\mu}_\gamma)^T\bm{w}}{\bm{w}^T(\bm{\Sigma}_\alpha+\bm{\Sigma}_\gamma)\bm{w}}
	\end{equation}
	Define $\mathbf{S}_W = \bm{\Sigma}_\alpha+\bm{\Sigma}_\gamma\quad \mathbf{S}_B=(\bm{\mu}_\alpha-\bm{\mu}_\gamma)(\bm{\mu}_\alpha-\bm{\mu}_\gamma)^T$, the target function can be rewritten as:
	\begin{equation}
	\label{eq:6}
	\bm{w}^* = \mathop{\text{arg~max}}\limits_{\bm{w}}~J(\bm{w}) = \mathop{\text{arg~max}}\limits_{\bm{w}}\dfrac{\bm{w}^T\mathbf{S}_B\bm{w}}{\bm{w}^T\mathbf{S}_W\bm{w}}
	\end{equation}
	The closed-form of the optimal solution of $\bm{w}$ can be solved by deriving equation \ref{eq:6} and making it zero $\nabla~J(\bm{w})=0$.
	\begin{equation}
	\label{eq:7}
	\bm{w}\propto\mathbf{S}_W^{-1}(\bm{\mu}_\alpha-\bm{\mu}_\gamma)
	\end{equation}
	Equation \ref{eq:7} is known as Fisher linear discriminant.
	\section{Results}
	\label{sec:4}
	\subsection{Results of CCM}
	In this section, the relationship between the performance of CCM and digitizer's sampling properties is researched quantitatively, and the results of CCM are used to confirm the derivation.
	\subsubsection{Derivation of the relationship between CCM performance and Sampling Properties}
	\label{section:4.1.1}
	In this section, we make use of the theory proposed by J.Cang \cite{RN27,RN271} et al. and ADC quantization noise theory to research how digitizer's sampling properties influence the performance of CCM. Refer to their research, the variance of the integral of CCM $\sigma_{Q_n}^2$ can be calculated from
	\begin{equation}
	\label{eq.8}
	\sigma_{Q_n}^2 = \dfrac{1}{2}\left(S_{x\_I}+\dfrac{V_{n\_ADC}^2}{A^2\cdot F_s/2}\right)t_n
	\end{equation}
	where $S_{x\_I}$ is the noise power spectral density of front-end electronics, $A$ is the gain of the preamplifier, $V_{n\_ADC}^2$ is the square of ADC's noise, $F_S$ is the sampling rate of ADC and $t_n$ is the integral length. $V_{n\_ADC}^2$ can be derivated from the ADC quantization noise theory as \cite{mt-001}
	\begin{equation}
	\label{eq.9}
	V_{n\_ADC}^2 = \dfrac{1}{12}\left(\dfrac{Full~Scale}{2^{\mathrm{ENOB}}}\right)^2
	\end{equation}
	Substituting equation \ref{eq.9} into equation \ref{eq.8}:
	\begin{align}
	\label{eq.10}
	\sigma_{Q_n}^2 &= \frac{1}{12}\cdot\left(\dfrac{Full~Scale}{ 2^{\mathrm{ENOB}}}\right)^2\cdot\frac{t_n}{F_S\cdot A^2} + \frac{1}{2}S_{x\_I}t_n\notag\\
	&= \left(\dfrac{Full~Scale}{\sqrt{F_S}\times 2^{\mathrm{ENOB}}}\right)^2 T_n +\sigma_{Q_{n\_I}}^2
	\end{align}
	where $T_n = \frac{t_n}{12A^2}$ and $\sigma_{Q_{n\_I}}^2=\frac{1}{2}S_{x\_I}t_n$. Then, the relationship between $\sigma_{CCM}^2$ and sampling properties ($Full Scale,~F_S$, ENOB) is derived by equation \ref{eq.10}. According to equation \ref{eq:1} and equation \ref{eq.10}, $\sigma_{CCM}^2$ is
	\begin{align}
	\label{eq:11}
	\sigma_{CCM}^2 &= \left(\dfrac{\partial CCM}{\partial Q_L}\right)^2\sigma_{Q_L}^2+\left(\dfrac{\partial CCM}{\partial Q_S}\right)\sigma_{Q_S}^2\notag\\
	&= \dfrac{1}{Q_L^4}\left[\left(\dfrac{Full~Scale}{\sqrt{F_S}\times 2^{\mathrm{ENOB}}}\right)^2\cdot(Q_S^2T_L+Q_L^2T_S)+Q_S^2\sigma_{Q_L\_I}^2+Q_L^2\sigma_{Q_S\_I}^2\right]
	\end{align}
	
	Moreover, the trend of FoM values can also be derived from above results:
	\begin{align}
	\label{eq:12}
		FoM &= \dfrac{|CCM_\alpha-CCM_\gamma|}{2.355\times(\sigma_{CCM-\alpha}+\sigma_{CCM-\gamma})}\notag\\
	&= \dfrac{\frac{1}{2.355}\times|CCM_\alpha-CCM_\gamma|}{\displaystyle\sum_{i\in\alpha,\gamma}\left[\dfrac{1}{Q_L^2}\sqrt{(Q_S^2T_L+Q_L^2T_S)\times\left(\dfrac{Full~Scale}{\sqrt{F_S}\cdot 2^{\mathrm{ENOB}}}\right)^2 + (Q_S^2\sigma_{Q_{L\_I}}^2+Q_L^2\sigma_{Q_{S\_I}}^2)}\right]_i}\notag\\
	&= \dfrac{1}{\sqrt{Ax+B}+\sqrt{Cx+D}}	
	\end{align}
    where $x = \left(\frac{Full~Scale}{\sqrt{F_S}\cdot2^{\mathrm{ENOB}}}\right)^2$, $A = \left(\frac{2.355}{|CCM_\alpha-CCM_\gamma|}\right)^2\left(\frac{Q_S^2T_L+Q_L^2T_S}{Q_L^4}\right)_\alpha$, $B = \left(\frac{2.355}{|CCM_\alpha-CCM_\gamma|}\right)^2\left(\frac{Q_S^2\sigma_{Q_{L\_I}}^2+Q_L^2\sigma_{Q_{S\_I}}^2}{Q_L^4}\right)_\alpha$, $C = \left(\frac{2.355}{|CCM_\alpha-CCM_\gamma|}\right)^2\left(\frac{Q_S^2T_L+Q_L^2T_S}{Q_L^4}\right)_\gamma$, $D = \left(\frac{2.355}{|CCM_\alpha-CCM_\gamma|}\right)^2\left(\frac{Q_S^2\sigma_{Q_{L\_I}}^2+Q_L^2\sigma_{Q_{S\_I}}^2}{Q_L^4}\right)_\gamma$. The $x$ values of the digitizers used in this paper are listed in Table \ref{tab:1}. Moreover, in this research, the fluctuation of $|CCM_\alpha-CCM_\gamma|$ is pretty small and the change of it has little impact on the parameters $A, B, C$ and $D$. If the waveforms are accurately aligned, these parameters can be regarded as constants.
	\subsubsection{Experimental Results}
	\begin{figure}[htbp]
		\centering 
		\includegraphics[height = 5.3cm]{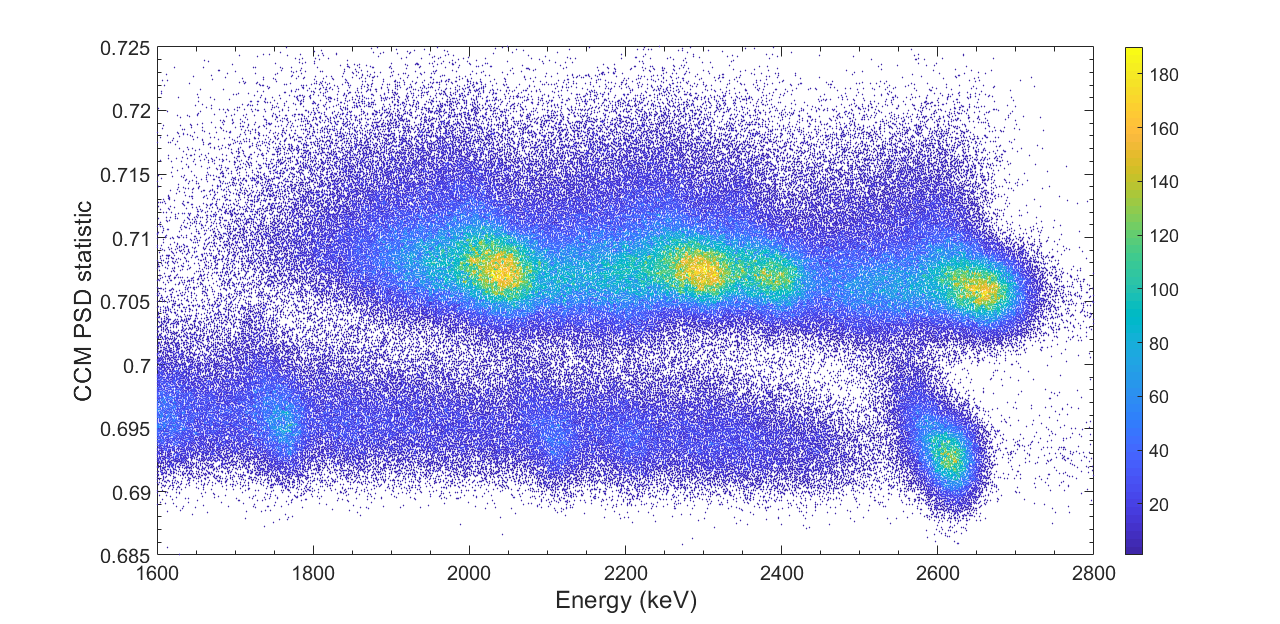}
		~
		\includegraphics[height = 5.3cm]{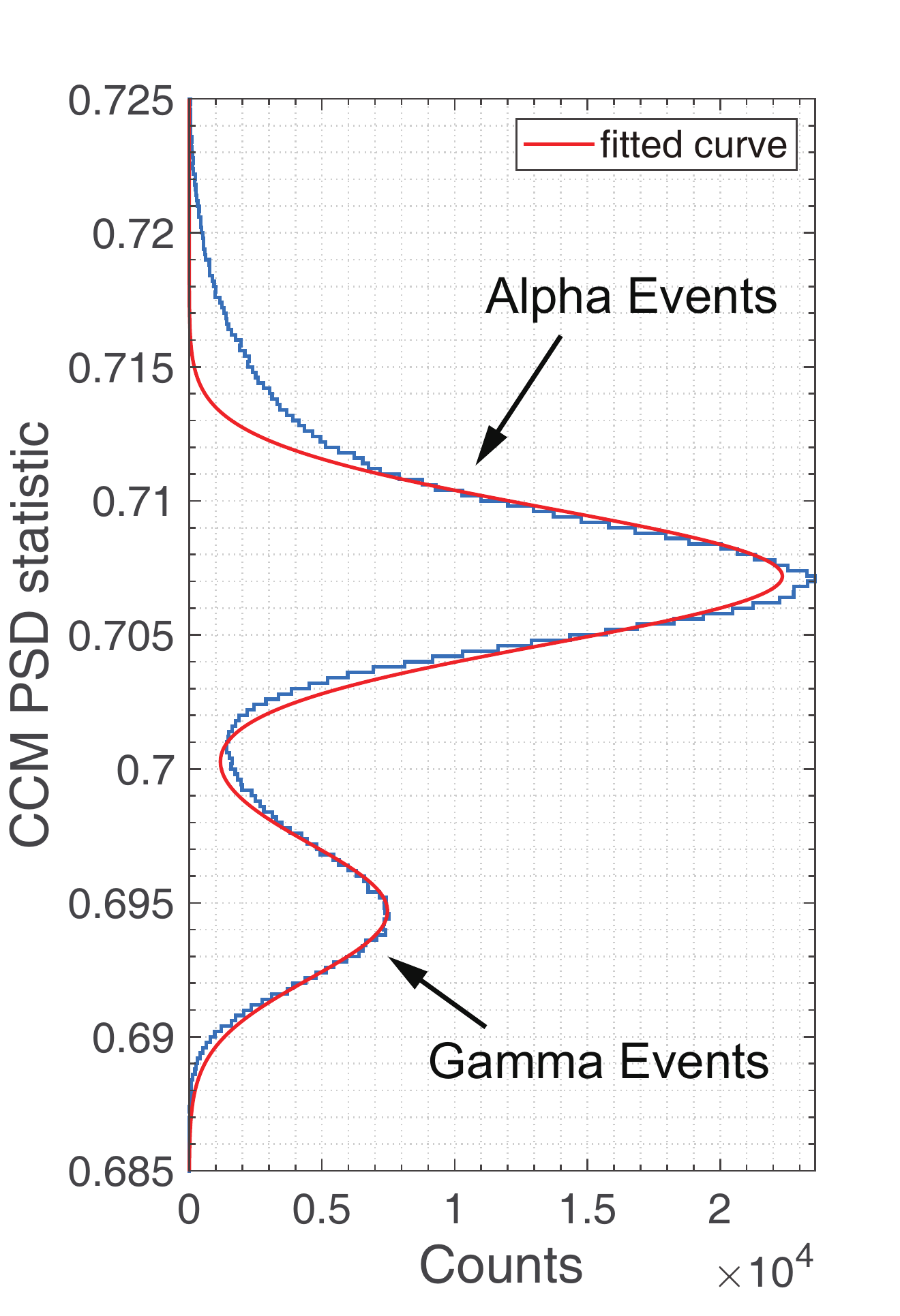}
		\caption{\label{fig:4} 2D distribution of CCM test statistic and energy measured by a 500 MSPS 12-Bit ISLA212P50 ADC and the figure includes 1,100,000 particle events.}
	\end{figure}
	The relationship between FoM values and sampling properties as equation \ref{eq:12} is evaluated by the experimental results of CCM in this section, and how to select the training datasets for LS and LDA algorithms is also introduced.
	
	CCM test statistics and energy are plotted as a 2D distribution figure with the $\gamma$-rays whose energy ranging from 1600 keV to 2800 keV and the $\alpha$ particles whose integral charges located in the same region. Moreover, the statistical histogram of corresponding CCM feature values is shown in figure \ref{fig:4}. To evaluate the $\alpha/\gamma$ discrimination efficiency, the figure-of-merit (FoM) is introduced, the values of it define the separation between $\alpha$ signals and $\gamma$ signals. The definition of FoM is:
	\begin{equation}
	\label{eq:13}
	\mathrm{FoM} = \dfrac{|\mathrm{Peak_\alpha-Peak_\gamma}|}{\mathrm{FWHM_\alpha+FWHM_\gamma}}
	\end{equation}
	
\begin{figure}[htbp]
		\centering 
		\includegraphics[width=0.8\textwidth,clip]{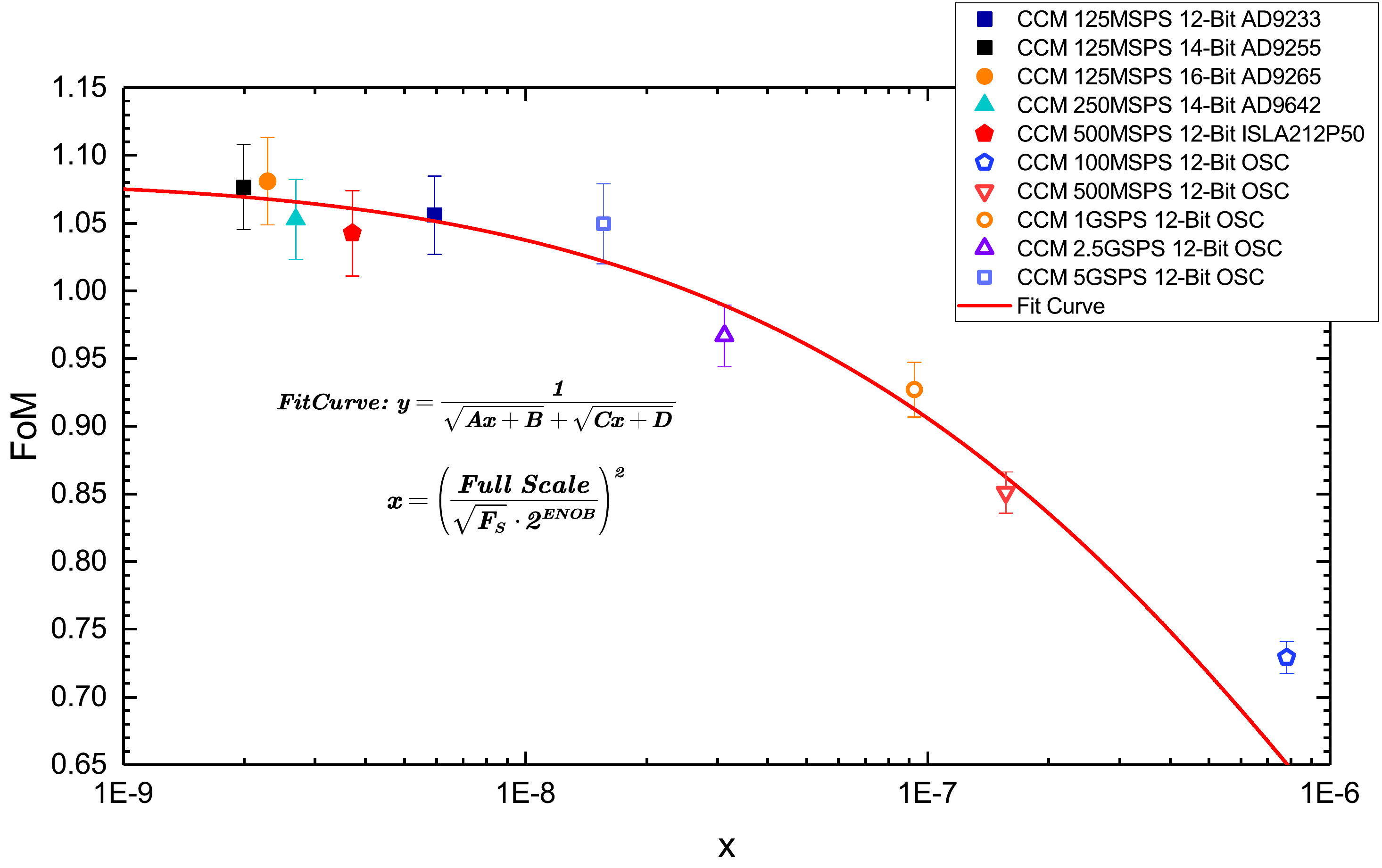}
		\caption{\label{fig:5} $\alpha/\gamma$ discrimination results of CCM and the corresponding fitting curve.}
	\end{figure}

	Figure \ref{fig:5} illustrates the relationship between the discrimination factor FoM and $x = \left(\frac{Full~Scale}{\sqrt{F_S}\cdot 2^{\mathrm{ENOB}}}\right)^2$ using the experimental results measured with different digitizers. The experimental results are quite close to the fit curve whose function form is  $1/(\sqrt{Ax+B}+\sqrt{Cx+D})$ as previously written, and the best FoM value of CCM can be predicted from the curve, which may achieve 1.1 as the value of x tends to zero.  Thus, we have verified the relationship derived from equation \ref{eq:12} in section \ref{section:4.1.1} and may help researchers to evaluate the PSD efficiency based on the digitizer's performance.
	
	\subsection{Results of LS \& LDA}
	\subsubsection{Experimental Results}
\begin{figure}[htbp]
		\centering 
		\includegraphics[width=0.64\textwidth,clip]{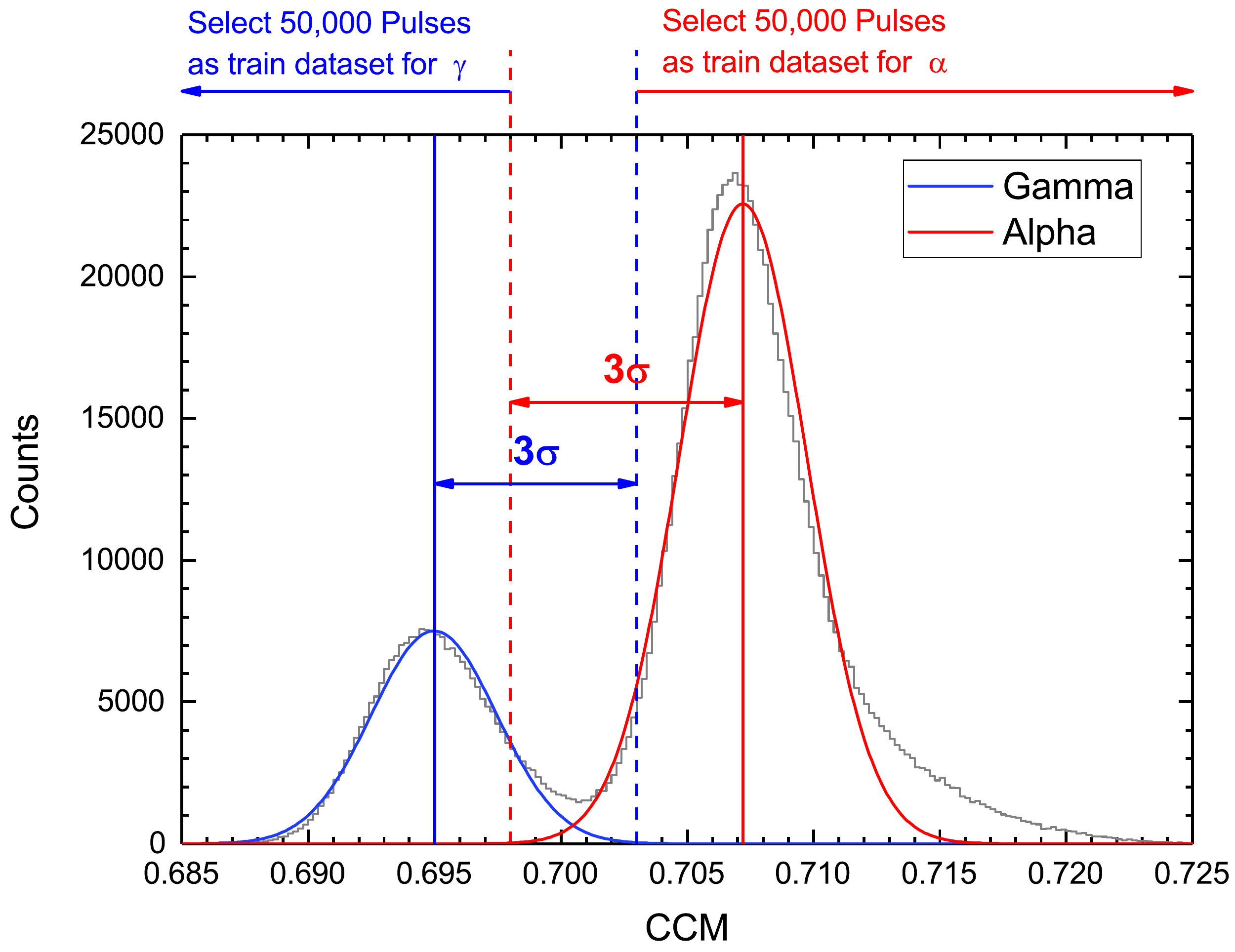}
		\caption{\label{fig:6} The $\alpha/\gamma$ signals' distribution with CCM algorithms, 50,000 waveforms each are selected as $\alpha/\gamma$ training datasets by 3$\sigma$ criterion.}
	\end{figure}

	From the 1.1 million waveforms of every digitizer's dataset, the training datasets of LS and LDA are selected based on the discrimination results of CCM, and figure \ref{fig:6} illustrates the selection strategy: using the statistical histogram of CCM feature values and choosing 50000 waveforms each as $\alpha/\gamma$ training datasets by 3$\sigma$ criterion. The other 1,000,000 waveforms of every dataset are selected as test dataset. 
	
\begin{figure}[htbp]
		\centering 
		\subfigure[]{
			\begin{minipage}[t]{0.49\linewidth}\label{fig:7a}
				\centering
				\includegraphics[width=\textwidth]{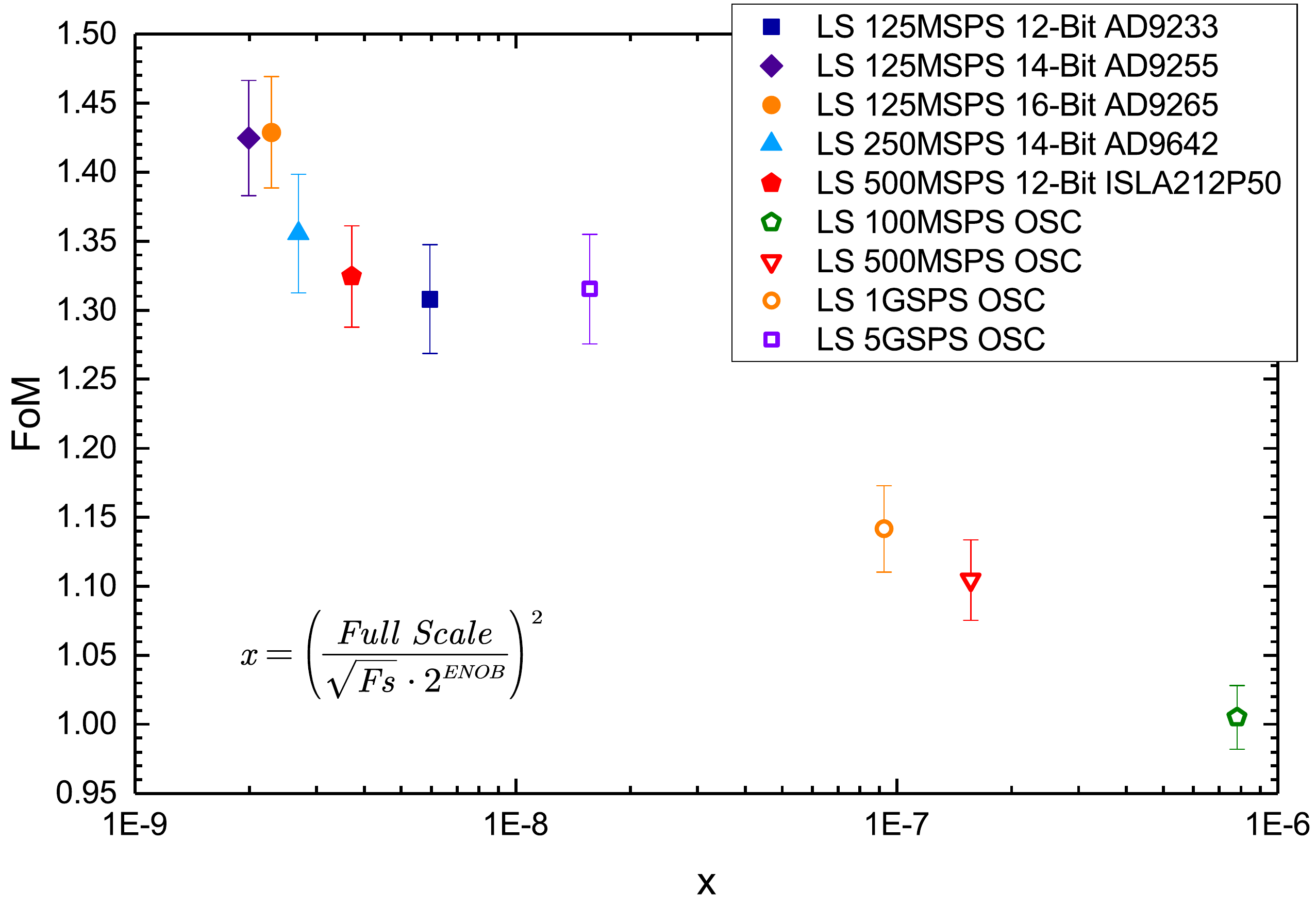}
		\end{minipage}}
		\subfigure[]{
			\begin{minipage}[t]{0.49\linewidth}\label{fig:7b}
				\centering
				\includegraphics[width=\textwidth]{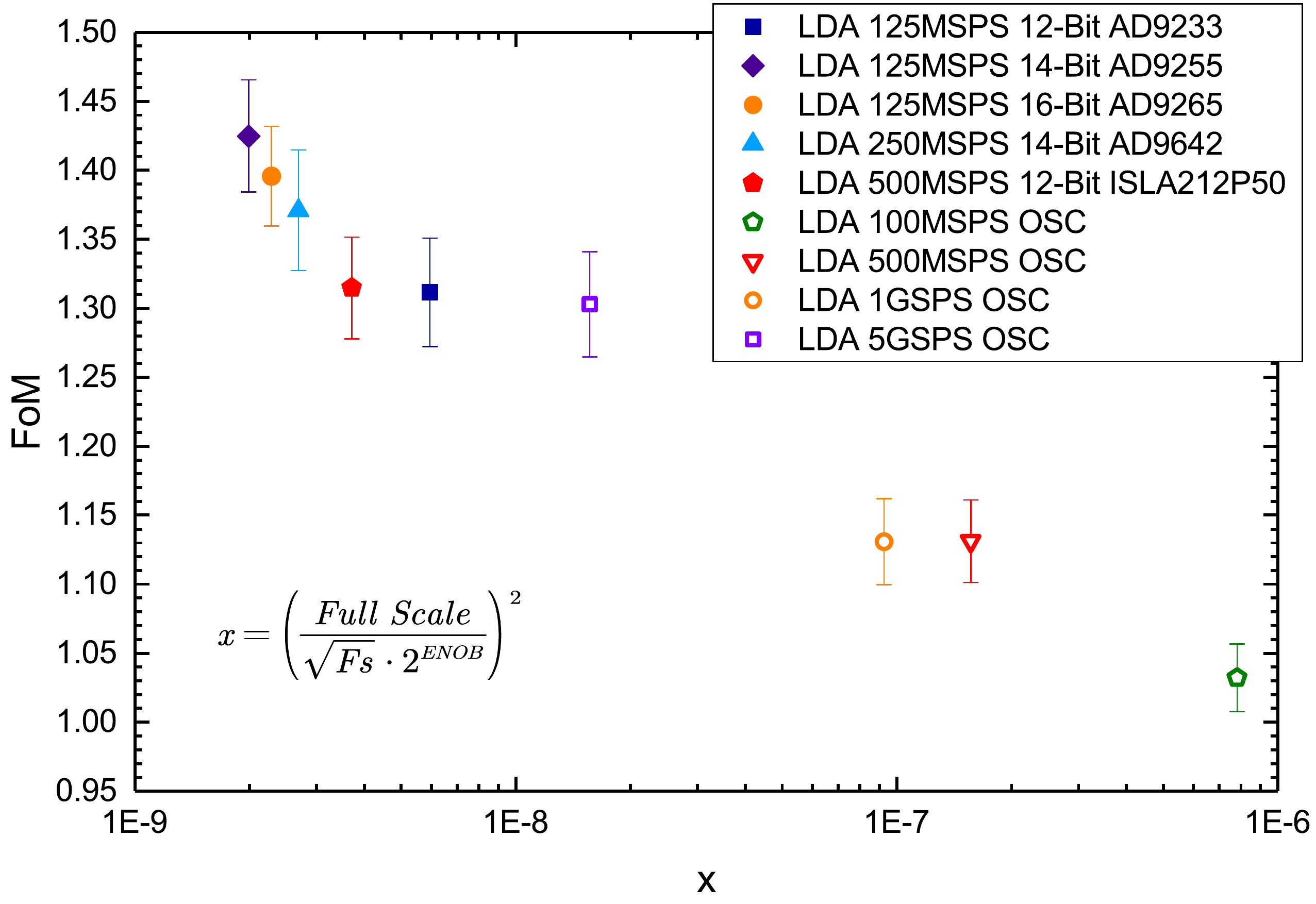}
		\end{minipage}}
		\caption{\label{fig:7} The FoM values changed with $x = \left(\frac{Full~Scale}{\sqrt{F_S}\cdot 2^{\mathrm{ENOB}}}\right)^2$ in LS/LDA algorithms}
	\end{figure}

	The relationship between the FoM obtained from LS/LDA and the sampling factor $x$ is given by figure \ref{fig:7a} and figure \ref{fig:7b}. Both algorithms achieved the best discrimination performances when using the 125 MSPS 14-Bit AD9255 ADC, and the FoM values are 1.408$\pm$0.039 (LS) and 1.425$\pm$0.040 (LDA). Although we have not developed the quantitative relationship between FoM values and sampling properties for the supervised discrimination algorithms like LS/LDA, the  value of $x = \left(\frac{Full~Scale}{\sqrt{F_S}\cdot 2^{\mathrm{ENOB}}}\right)^2$ still can provide some valuable significances. As shown in figure \ref{fig:7}, the FoM values show a decreasing trend with the increase of $x$.
	
	\subsubsection{The Influence of Waveform Recording Time on the Discrimination Results}
	\label{sec:4.2.2}
	By analyzing the relationship between waveform recording time and discrimination efficiency, we have a better understanding on how the sampling rate and ENOB of ADC influence the PSD results and provide some advice for researchers on how to choose the best waveform sampling intervals.
	\begin{figure}[h]
		\centering 
		\includegraphics[width=0.8\textwidth,clip]{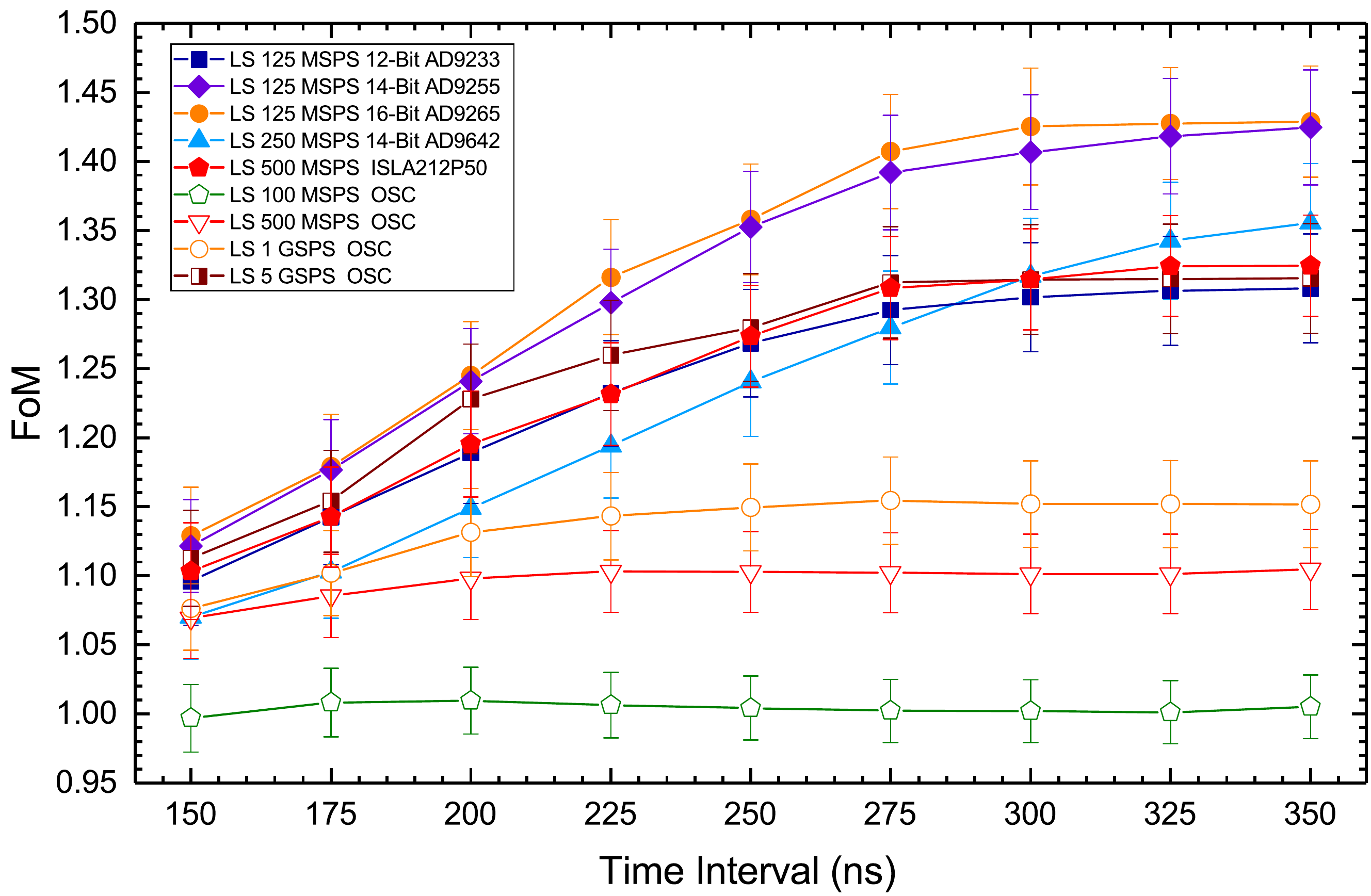}
		\caption{\label{fig:8}The relationship between FoM values and the recording time of waveforms. (The start point of each waveform is 0 ns as figure \ref{fig:2}, and the endpoint ranges from 150 ns to 350 ns.)}
	\end{figure}

	Figure \ref{fig:8} illustrated how the FoM values changed with the waveform recording time measured by different digitizers. The left edge (starting point) of each waveform is 0 ns (as shown in figure \ref{fig:2}), and the right edge ranges from 150 ns to 350 ns. For the high ENOB digitizers (AD9233, AD9255, AD9265, AD9642 and ISLA212P50) whose ENOB > 10, the discrimination performance is significantly improved as the increase of waveform recording time. However, for the low ENOB digitizers (LeCroy oscilloscope whose ENOB is 7.5), the longer waveform recording time has almost no impact on the discrimination efficiency except for the results from 5 GSPS dataset. The phenomenon can be analyzed from two perspectives:
	\begin{itemize}
		\item If we use a digitizer with low ENOB to sample the waveform, the ADC quantization noise will be similar to or higher than the height of waveforms in 150 $\sim$ 350 ns. So the information of the waveform tail cannot be accurately measured by the oscilloscope with ENOB = 7.5. However, a digitizer with high ENOB can avoid this problem and improve the discrimination performance by increasing the recording time of waveforms.   
		\item Due to the high sampling rate, the discrimination results from the 5 GSPS oscilloscope dataset are also improved with the increase of waveforms' length, because the oversampling compensates the information loss caused by the quantization noise\cite{AN118}.
	\end{itemize}
	To verify the discussion results, we select the average waveforms from a AD9642 ADC (250 MSPS, ENOB = 11.06), and calculate the  difference between the $\alpha/\gamma$ waveforms as eq.\ref{eq:14} to quantitatively analyze the requirement for the digitizer's ENOB.
	\begin{equation}
	\label{eq:14}
	\Delta = \dfrac{\mathrm{Mean}_\alpha-\mathrm{Mean}_\gamma}{\mathrm{Max}_{\mathrm{Mean}_\gamma}}
	\end{equation}
	where $\mathrm{Mean_\alpha}$ and $\mathrm{Mean_\gamma}$ represent the average waveforms of $\alpha/\gamma$ which shown in figure 2, and $\mathrm{Max_{Mean_\gamma}}$ is the max value of the average $\gamma$ waveform.

\begin{figure}[h]
		\centering 
		\includegraphics[width=0.8\textwidth,clip]{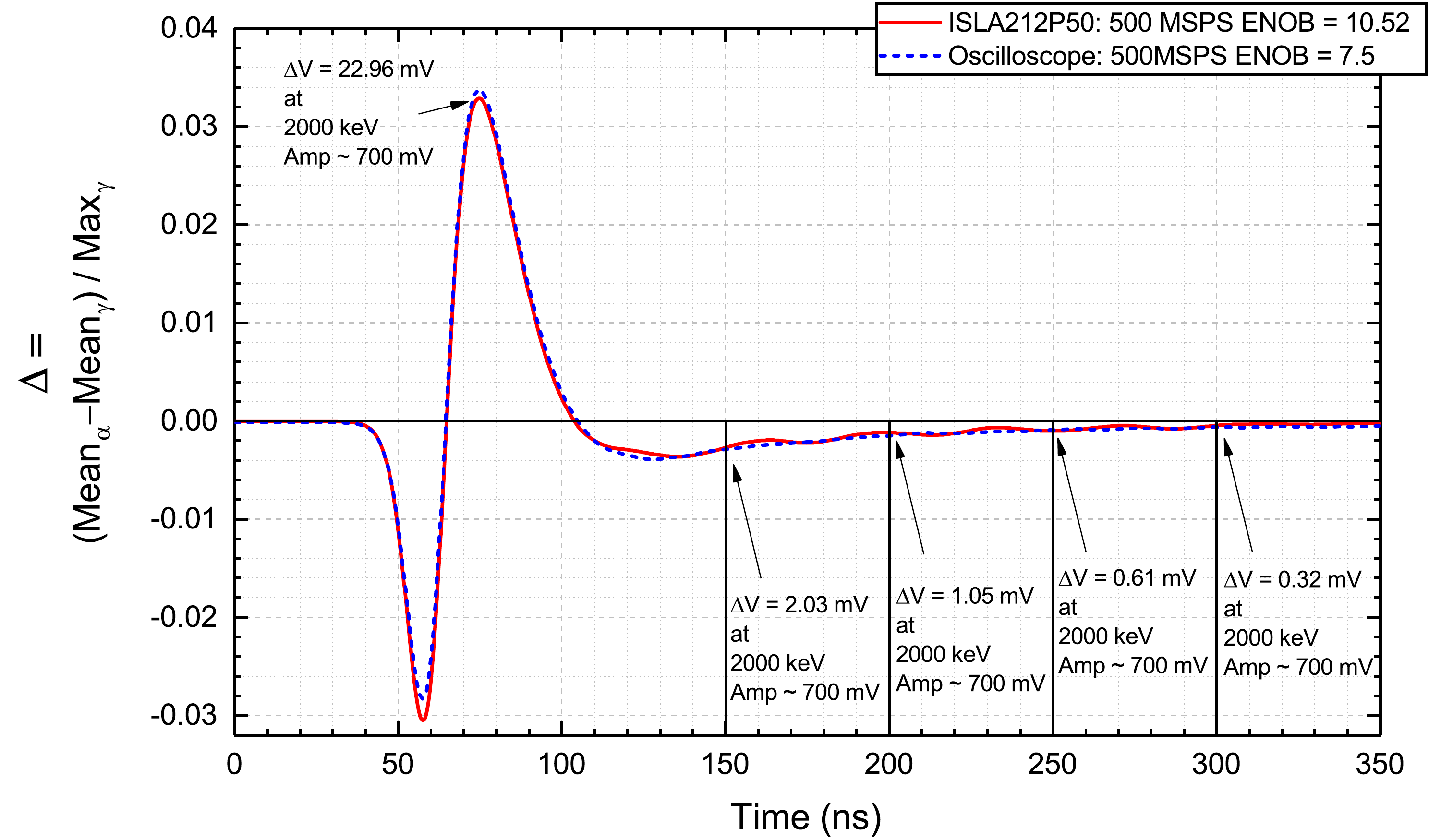}
		\caption{\label{fig:ext}The difference between average $\alpha$ and average $\gamma$ waveforms from ISLA212P50 ADC and LeCroy oscilloscope. Corresponding height differences $\Delta V$ at 150/200/250/300 ns are marked on the figure.}
	\end{figure}

	The result is shown in figure \ref{fig:ext}. For a $\gamma$ signal whose energy is 2000 keV, the maximum amplitude is 700 mV, then the voltage difference between $\alpha/\gamma$ can be calculated from $\Delta V = \Delta\times700~\mathrm{mV}$. The corresponding $\Delta V$ values at 150/200/250 ns are 2.03/1.05/0.61 mV. Moreover, the minimum effective resolutions of the digitizers used in this research are listed in Table \ref{tab:2}. Comparing the minimus resolutions of digitizers and $\Delta V$, it can be concluded that a digitizer with high ENOB can effectively distinguish the nuances of the $\alpha/\gamma$ waveforms in the range of 150 ns to 350 ns. 
	\begin{table}[htbp]
		\centering
		\caption{\label{tab:2} Minimus resolutions ($\mathrm{Full~Scale}/2^{\mathrm{ENOB}}$) of digitizers}
		\smallskip
		\begin{tabular}{|l|c|c|c|c|c|c|}
			\hline
			ADC&AD9233 & AD9255 &AD9265 & AD9642 & ISLA212P50 & Oscilloscope \\ 
			\hline 
			Full Scale & 2000 mV & 2000 mV & 2000 mV & 1750 mV & 2000 mV & 1600 mV\\
			ENOB & 11.18 & 11.97 & 11.87 & 11.06 & 10.52 & 7.5\\
			$\mathrm{Full~Scale}/2^{\mathrm{ENOB}}$ & 0.86 mV & 0.50 mV & 0.53 mV & 0.82 mV & 1.36 mV & 8.84 mV\\
			\hline
		\end{tabular}
	\end{table}
	
	Based on the analysis, a conclusion can be drawn that we should extend the sampling length enough to keep more tail information of waveforms in the experiment and use a digitizer with higher ENOB to distinguish the detail information. 
	
	\section{Discussion}
	\subsection{Comparison of the Algorithms}
	The discrimination results of three algorithms are drawn in one figure to compare their similarities and differences. As shown in figure 9, the LS and LDA have similar discrimination capability, and their discrimination efficiency is much better than CCM especially in the case of using a digitizer with the low sampling rate and low ENOB. The best FoM values for the algorithms are 1.429$\pm$0.040 (LS), 1.425$\pm$0.041 (LDA) and 1.081$\pm$0.032 (CCM). Moreover, the FoM value of LS/LDA is 1.006$\pm$0.023/1.032$\pm$0.025 and 0.729$\pm$0.012 for CCM when using the data from a 100 MSPS, ENOB = 7.5 oscilloscope.
	
	CCM is an algorithm based on charge comparison; therefore some nuances might be ignored due to the information compression of the integration.  Meanwhile, according to the description in section \ref{sec:4.2.2}, the LS and LDA algorithms can more effectively utilize the information of waveforms' tail and distinguish even the small differences which are under the sensitiveness of the CCM method. 
	\begin{figure}[htbp]
		\centering 
		\includegraphics[width=0.9\textwidth]{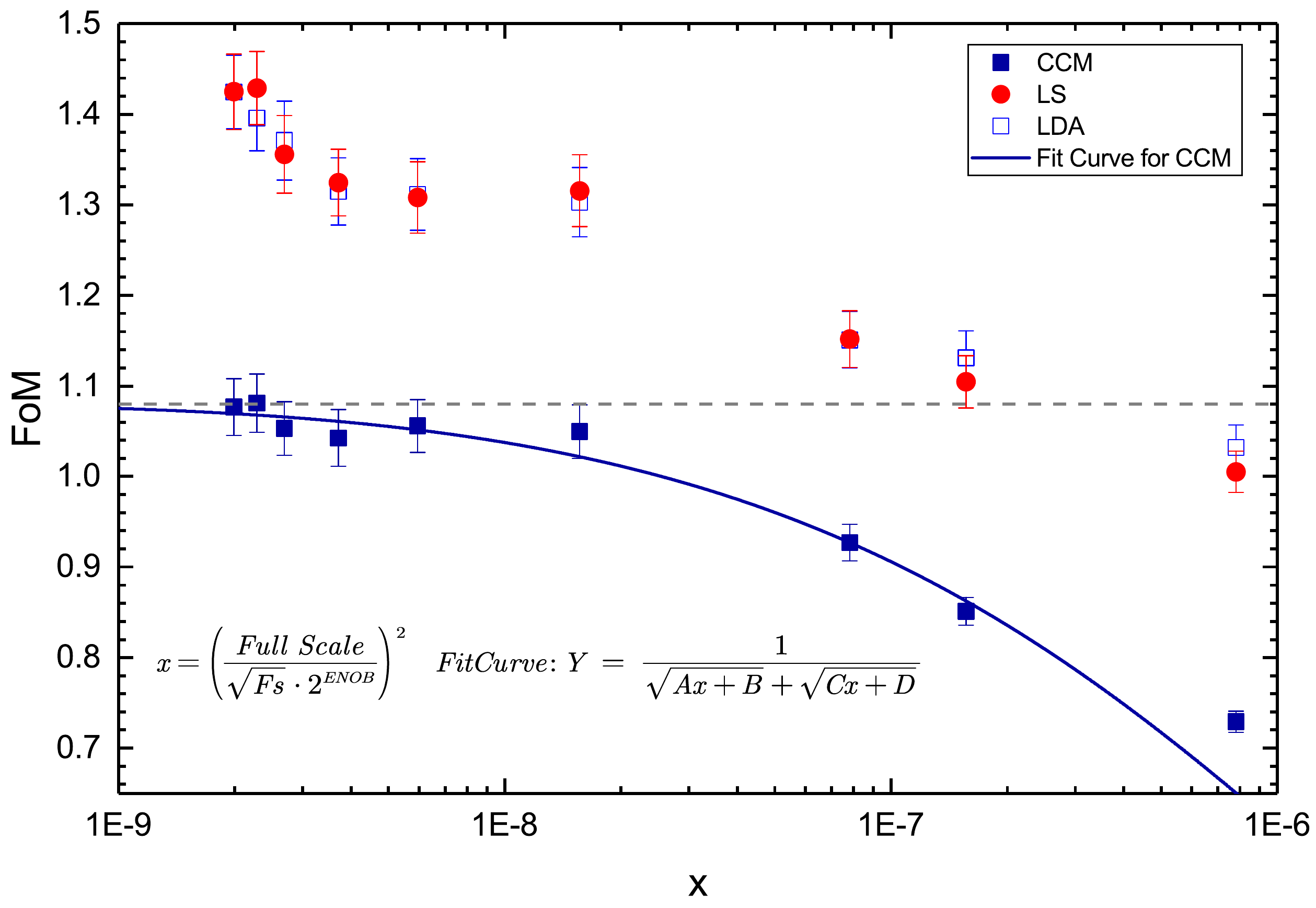}
		\caption{\label{fig:9} Comparison the FoM values of CCM, LS and LDA.}
	\end{figure}
	
	\subsection{LaBr3:Ce PSD and Linear Classification Algorithm}
	According to the improvement of the discriminating results by using the LS/LDA algorithm and the dimensionality reduction results from t-SNE in section \ref{sec:3.2}, we can speculate that the $\alpha/\gamma$ waveforms of the $\mathrm{LaBr_3}$:Ce scintillation detectors are approximately linearly separable and the original distribution is close to a Gaussian distribution. From the perspective of the light production in scintillator detectors and the electronics generated in the PMT, the $\alpha/\gamma$ signals can be regarded as stationary random signals. Moreover, since each point on the waveform is a superposition of a large number of electronics, the random parameters of signals should approximately obey the Gaussian distribution. Therefore, it can be further inferred that the PSD algorithms based on linear model can also provide good discrimination results in other scintillation detectors.
	
	Although some nonlinear classifiers such as soft interval support vector machine (SVM), decision tree and neural network may provide better performance on the discrimination of $\alpha/\gamma$signals, the PSD algorithms based on linear model still have considerable advantages. The linear PSD algorithms  have less demand for computation and hardware, so it's easier to achieve the target of real-time deployment with them, such as their implemented in FPGA. Moreover, linear discrimination algorithms reduce the need for complex readout electronics and then reduce the cost of the data acquisition and data storage system. 
	
	\section{Conclusion}
	This paper analyzes the performance of three linear classification algorithms (CCM, LS and LDA) on $\alpha/\gamma$ discrimination for $\mathrm{LaBr_3}$:Ce scintillation detectors. Moreover, several digitizers are used to research the relationship between the sampling properties and discrimination efficiency. The classification results of LS are similar to LDA, and their performance is much better than CCM. The FoM factors of LS/LDA is 1.424$\pm$0.042/1.425$\pm$0.041 with AD9255 (125 MSPS, 14-Bit, ENOB = 11.97), and in the same case, the FoM of CCM is 1.077$\pm$0.031. Besides, the LS/LDA algorithms can also provide optimal discrimination results when using a low sampling rate and ENOB digitizer.
	
	In addition, through the relationship between sampling properties and classification results, we find that ENOB plays a more important role in the improvement of discrimination efficiency than the sampling rate.  Moreover, the high ENOB digitizer systems are able to exploit the minimal differences found in the tail of the waveforms so to enhance the discrimination capability of the system, by using longer integration intervals, without being affected by the electronic noise. The discrimination capability of a digitizer with enough sampling rate and high ENOB (e.g. AD9233 125 MSPS 12-Bit ENOB = 11.18) is much better than the digitizer with very high sampling rate but low ENOB (e.g. Oscilloscope 1 GSPS ENOB = 7.5). Moreover, the FoM factors of the three methods will decrease as the factor $x=\left(\frac{Full~Scale}{\sqrt{F_S}\cdot 2^{\mathrm{ENOB}}}\right)^2$ is increasing. This relationship can help researchers to make a trade-off between sampling properties and desirable discrimination results.
	
	The LS/LDA algorithms require fewer hardware resources for online deployment than nonlinear classification algorithms like neural-network, and their discrimination capabilities are much better than the traditional PSD methods like CCM. Therefore,  the implementation of LS/LDA algorithms may be helpful for system developers to reduce the demand for data acquisition systems, data processing systems and data storage systems, thereby reducing the cost of the electronics.

	\acknowledgments
	
	This work is supported by the National Key Research and Development Program of China 
	
	\noindent{(2017YFA0402202)}.
	
	 We would like to thank those who collaborated on the CDEX, and also thankwords of deep appreciation go to Professor Yulan Li, Qian Yue, Litao Yang and Guang Meng for their invaluable advices, supports and various discussions over the years at the Tsinghua University DEP (Department of Engineering Physics).
	 
	  We are grateful for the patient help of Yu Xue, Wenping Xue, and Jianfeng Zhang. They are seasoned, full-stack hardware technologists with rich experience of solder and rework in the electronics workshop at DEP.

	


\begin{thebibliography}{99}
		
		
		
		\bibitem{dorenbos2002light}
		P. Dorenbos, \emph{Light output and energy resolution of Ce$^{3+}$-doped scintillators}, \href{https://sciencedirect.xilesou.top/science/article/pii/S0168900202007040}{\emph{Nuclear Instruments and Methods A} {\bf 486} (2002) 208-213}.
		
		\bibitem{RN24}
		M. Flaska, M. Faisal, D. D. Wentzloff and S. A. Pozzi, \emph{Influence of sampling properties of fast-waveform digitizers on neutron-gamma-ray, pulse shape discrimination for orgranic scintillation detectors}, \href{https://sciencedirect.xilesou.top/science/article/pii/S0168900213009844}{\emph{Nuclear Instruments and Methods A} {\bf 729} (2013) 456-462}.
		
		\bibitem{RN22}
		D. Cester, M. Lunardon, G. Nebbia, L. Stevanato, G. Viesti, S. Petrucci et al., \emph{Pulse shape discrimination with fast digitizers}, \href{https://sciencedirect.xilesou.top/science/article/pii/S016890021400196X}{\emph{Nuclear Instruments and Methods A} {\bf 748} (2014) 33-38}.
		
		\bibitem{ref4}
		L. Bardelli L, G. Poggi, \emph{Digital-sampling systems in high-resolution and wide dynamic-range energy measurements: Comparison with peak sensing ADCs}, \href{http://www.lnl.infn.it/~garfweb/e_digit/private_docs/references/bardelli_poggi_Digital-sampling_systems_comparison_with_peak_sensing_ADCs_NIMA_560_2006_517.pdf}{\emph{Nuclear Instruments and Methods A} {\bf 560} (2006) 517-523}.
		
		\bibitem{RN25}
		J. Zhang, M. E. Moore, Z. Wang, Z. Rong, C. Yang and J. P. Hayward, \emph{Study of sampling rate influence on neutron-gamma discrimination with stilbene coupled to a silicon photomultiplier}, \href{https://sciencedirect.xilesou.top/science/article/abs/pii/S0969804317300787}{\emph{Appl Radiat Isot.} {\bf 128} (2017) 120-124}.
		
		
		\bibitem{RN27}
		J. Cang, T. Xue, M. Zeng, Z. Zeng, H. Ma, J. Cheng et al., \emph{Optimal design of waveform digitisers for both energy resolution and pulse shape discrimination}, \href{https://arxiv.xilesou.top/abs/1712.05207}{\emph{Nuclear Instruments and Methods A} {\bf 888} (2018) 96-102}.
		
		\bibitem{RN271}
		T. Xue, J. Zhu, J. Wen, J. Cang, Z. Zeng, L. Wei et al., \emph{Optimization of Energy Resolution and Pulse Shape Discrimination for a CLYC Detector with Integrated Digitizers}, \href{https://arxiv.org/abs/1911.10042}{\emph{arXiv:}1911.10042}.
		
		\bibitem{RN28}
		J. Zhu, G. Gong, T. Xue, Z. Cao, L. Wei and J. Li, \emph{Preliminary Design of Integrated Digitizer Base for Photomultiplier Tube}, \href{https://arxiv.xilesou.top/abs/1806.08937}{\emph{IEEE Transactions on Nuclear Science} {\bf 66} (2019) 1130-1137}.
		
		\bibitem{maaten2008visualizing}
		L. Maaten and G. Hinton, \emph{Visualizing data using t-SNE}, \href{http://www.jmlr.org/papers/v9/vandermaaten08a.html}{\emph{Journal of machine learning research} {\bf 9} (2008) 2579-2605}.
		
		\bibitem{RN29}
		M. Zeng, J. Cang, Z. Zeng, X. Yue, J. Cheng, Y. Liu et al., \emph{Quantitative analysis and efficiency study of PSD methods for a $\mathrm{LaBr_3}$:Ce detector}, \href{https://arxiv.xilesou.top/abs/1504.05346}{\emph{Nuclear Instruments and Methods A} {\bf 813} (2016) 56-61}.
		
		\bibitem{bishop2006pattern}
		C. M. Bishop, \emph{Pattern recognition and machine learning}. Springer Science+ Business Media, 2006. 
		
		\bibitem{fisher1936use}
		R. A. Fisher, \emph{The use of multiple measurements in taxonomic problems}, \href{https://onlinelibrary_wiley.xilesou.top/doi/abs/10.1111/j.1469-1809.1936.tb02137.x}{\emph{Annals of eugenics} {\bf 7} (1936) 179-188}.
		
		\bibitem{mt-001}
		W. Kester, \emph{Taking the mastery out of the infamous formula, "SNR = 6.02N + 1.76 dB," and why you should care}, \href{https://www.analog.com/media/en/training-seminars/tutorials/MT-001.pdf}{Analog Device MT-001}.
	
	    \bibitem{AN118}
	    Silicon Laboratories, \emph{Improving ADC resolution by oversampling and averaging}, \href{https://www.silabs.com/documents/public/application-notes/an118.pdf}{Silicon Labs AN118}.
		
		
		
		
		
		
	\end{thebibliography}
\end{document}